\documentclass[aps,prl,twocolumn,superscriptaddress,floatfix]{revtex4-2}

\usepackage{graphicx}
\usepackage{amsmath,amssymb}
\usepackage{hyperref}
 \usepackage{cleveref} 
\usepackage{times}
\usepackage{subcaption}
\usepackage{color}

\begin{document}

\title{Competing Length Scales and Symmetry Frustration Govern Non-Universal Melting in 2D Core-softened Colloidal Crystals}

\author{Thiago Puccinelli}
\affiliation{Instituto de F\'isica, Universidade de S\~ao Paulo (USP), 05508-090, S\~ao Paulo, Brasil}

\author{Alexandre V. Ilha}
\affiliation{Departamento de F\'isica, Instituto de F\'isica e Matem\'atica, Universidade Federal de Pelotas, 96001-970, Pelotas-RS, Brasil}

\author{Jos\'e Rafael Bordin}
\email{jrbordin@ufpel.edu.br}
\affiliation{Departamento de F\'isica, Instituto de F\'isica e Matem\'atica, Universidade Federal de Pelotas, 96001-970, Pelotas-RS, Brasil}

\begin{abstract}
We investigate the melting behavior of two-dimensional colloidal crystals stabilized by a core-softened potential featuring two competing interaction length scales. Using molecular dynamics simulations, we analyze three polymorphic solid phases—low-density triangular, stripe, and kagome—and uncover distinct melting pathways. The triangular and kagome crystals undergo abrupt first-order transitions, driven by the interplay between energetic frustration and structural reorganization. In particular, the LDT phase melts through a sharp transition induced by a crossover between the two characteristic length scales. In contrast, the stripe phase exhibits a continuous transition with liquid-crystalline features: orientational and translational order decay gradually, while intra-stripe mobility persists, consistent with a KTHNY-like scenario. These findings demonstrate that melting in 2D soft-matter systems is inherently non-universal and governed by the competition between lattice symmetry, frustration, and multiple interaction scales. Our results provide microscopic insight into melting mechanisms beyond classical universality classes and offer guiding principles for the design of self-assembled materials with tunable phase behavior.

\end{abstract}

\maketitle

Melting in two-dimensional (2D) systems is often described by the Kosterlitz-Thouless-Halperin-Nelson-Young (KTHNY) theory~\cite{Mermin1968, Kosterlitz1973, Halperin1978, Nelson1979}, which predicts two continuous transitions: a solid transforms into a hexatic phase via dislocation unbinding, followed by a transition to an isotropic fluid through disclination unbinding. This framework has been validated in several colloidal and soft-matter systems with hexagonal symmetry and short-range repulsion~\cite{Strandburg1988, Zahn1999, Marcus1996, Quinn2001, Royall2024, Kosterlitz2016}. For instance, Royall et al.~\cite{Royall2024} showed how quasi-2D hard-sphere colloids exhibit melting behavior sensitive to polydispersity and confinement. Kosterlitz~\cite{Kosterlitz2016} revisited topological transitions and the broader applicability of defect-mediated melting.

However, deviations from the KTHNY scenario have emerged in systems with quenched disorder~\cite{Anderson2017}, anisotropic interactions~\cite{Tsiok2015, MassanaCid2024}, confinement~\cite{Hua2024}, or competing interactions~\cite{Prestipino2012, Fomin2021, MendozaCoto2024}. In these cases, melting can occur through a first-order transition without an intermediate hexatic, or involve mechanisms such as grain-boundary proliferation~\cite{OlsonReichhardt2010, Kapfer2015}. Kapfer and Krauth~\cite{Kapfer2015} mapped melting scenarios for soft and Yukawa disks, revealing a crossover from first-order to continuous transitions depending on interaction range.

These results challenge the universality of 2D melting and motivate further investigation in systems with structural complexity, frustration, or polymorphism~\cite{Charbonneau2024, Russo2018}. 

Core-softened (CS) potentials—comprising a repulsive core and a softer shoulder—coarse-grain a variety of soft-matter systems with two competing interaction lengths, including polymer-grafted nanoparticles, microgels, dendrimers, and screened colloids~\cite{Likos2001, Malescio2003, AngiolettiUberti2012, Ramos2011}. Experimental evidence for such interactions comes from DNA-coated colloids~\cite{Rogers2011}, hydrogels~\cite{Likos06}, and star polymers~\cite{Likos2002}. These materials often exhibit polymorphism, clustering, and reentrant melting. The presence of multiple scales stabilizes unconventional phases—such as stripes, honeycomb, and kagome lattices~\cite{Pieros2016, Lindquist2016, Cardoso2021, Nogueira2022, Nogueira2023}—making CS systems ideal for probing non-classical melting. However, the role of symmetry and frustration in these transitions remains poorly understood, despite its relevance for thermodynamic responses and design of programmable soft materials.

Here, we employ molecular dynamics simulations to investigate the melting behavior of three polymorphic two-dimensional phases stabilized by a core-softened (CS) potential: a low-density triangular lattice, a stripe phase, and a kagome crystal. We find that the triangular and kagome phases undergo sharp first-order melting transitions, whereas the stripe phase follows a continuous melting pathway consistent with the Kosterlitz-Thouless-Halperin-Nelson-Young (KTHNY) mechanism. These results demonstrate that melting in 2D polymorphic systems is inherently non-universal and governed by a subtle interplay between symmetry and frustration.

The system consists of \( N \approx 100000 \) particles confined to two dimensions and interacting via a CS pair potential featuring two characteristic length scales. The interaction combines a Lennard-Jones (LJ) core with an additional Gaussian repulsion,
$U_{CS}(r) = 4\epsilon \left[ \left( \frac{\sigma}{r} \right)^{12} - \left( \frac{\sigma}{r} \right)^6 \right] + u_0 \exp\left[ -\frac{1}{c_0^2} \left( \frac{r - r_0}{\sigma} \right)^2 \right],
\label{eq:CS}
$ with parameters \( u_0 = 5\epsilon \), \( c_0^2 = 1.0 \), and \( r_0 = 0.7\sigma \)~\cite{deOliveira2006, deOliveira2006b}. This ramp-like potential introduces a short-range shoulder at \( r_1 \approx 1.2\sigma \), associated with a local minimum in the interparticle force, and a longer-range scale at \( r_2 \approx 2.0\sigma \), related to a minimum in the fraction of imaginary modes from instantaneous normal mode analysis~\cite{BarrosdeOliveira2010}. The competition between these length scales stabilizes multiple crystalline phases and gives rise to water-like anomalies in both confined and bulk 2D and 3D systems~\cite{Bordin2018, Cardoso2021, Bordin2023}.

All quantities are reported in standard reduced Lennard-Jones units~\cite{allen2017}. Computational details, along with the definitions and behavior of thermodynamic, dynamic, and structural observables, are provided in Section I of the Supplementary Material. All input files, including simulation scripts, force field parameters, and initial configurations for each phase, are openly available in our GitHub repository: \url{https://github.com/thiagopuccinelli/2D-simulations-TSK}, enabling full reproducibility of the results presented in this work.

\begin{figure}[h!]
    \centering
    \begin{subfigure}[b]{0.2\textwidth}
        \includegraphics[width=\textwidth]{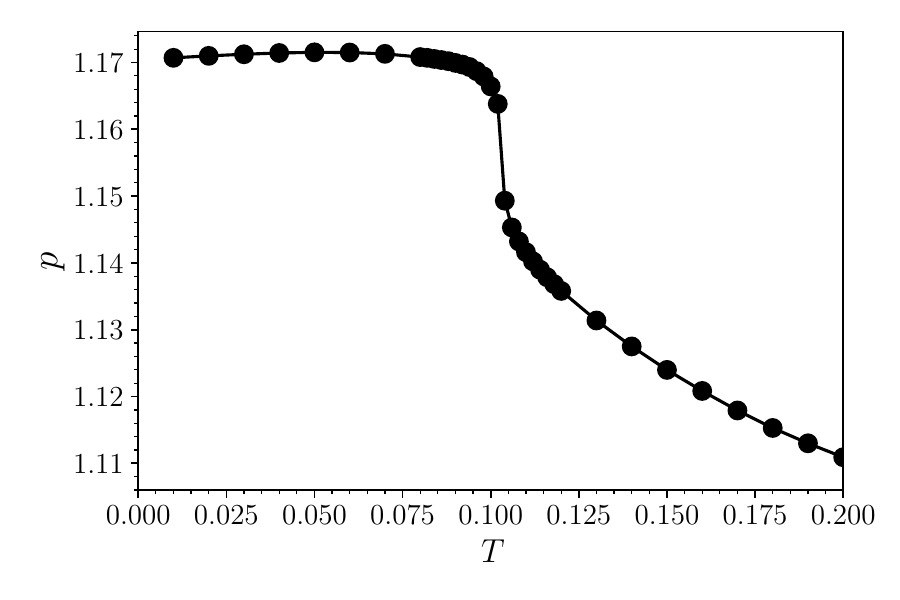}
        \caption{}
    \end{subfigure}
        \begin{subfigure}[b]{0.2\textwidth}
        \includegraphics[width=\textwidth]{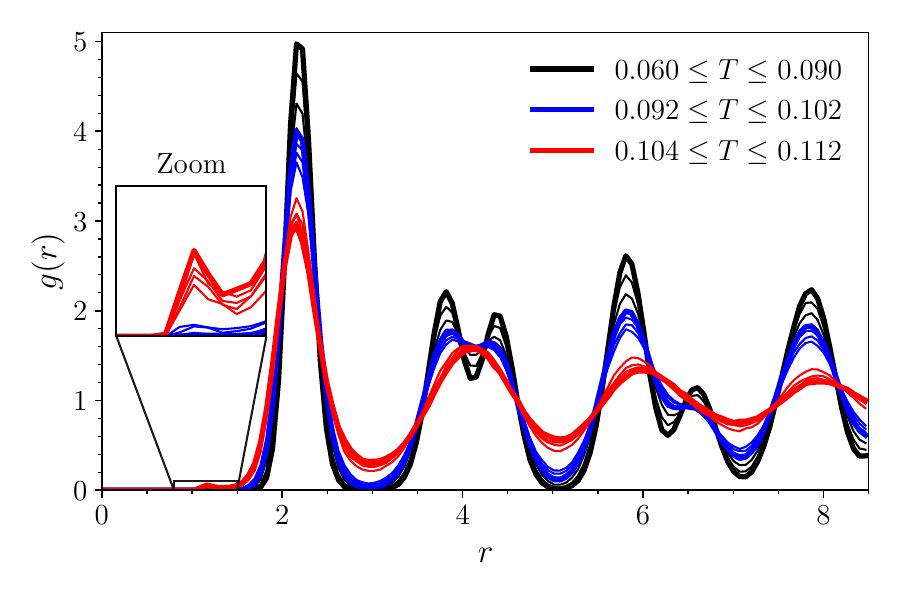}
        \caption{}
    \end{subfigure}
        \begin{subfigure}[b]{0.2\textwidth}
        \includegraphics[width=\textwidth]{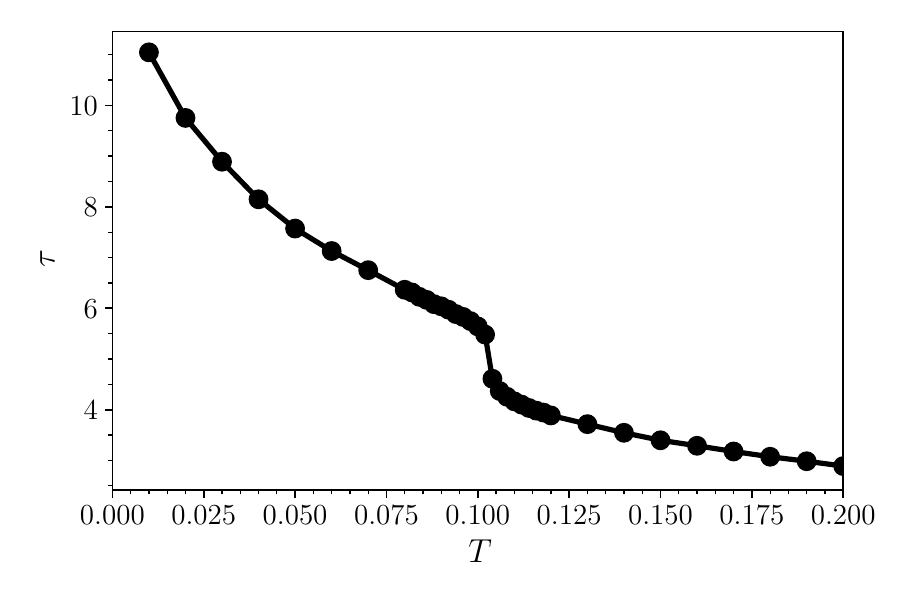}
        \caption{}
    \end{subfigure}
        \begin{subfigure}[b]{0.2\textwidth}
        \includegraphics[width=\textwidth]{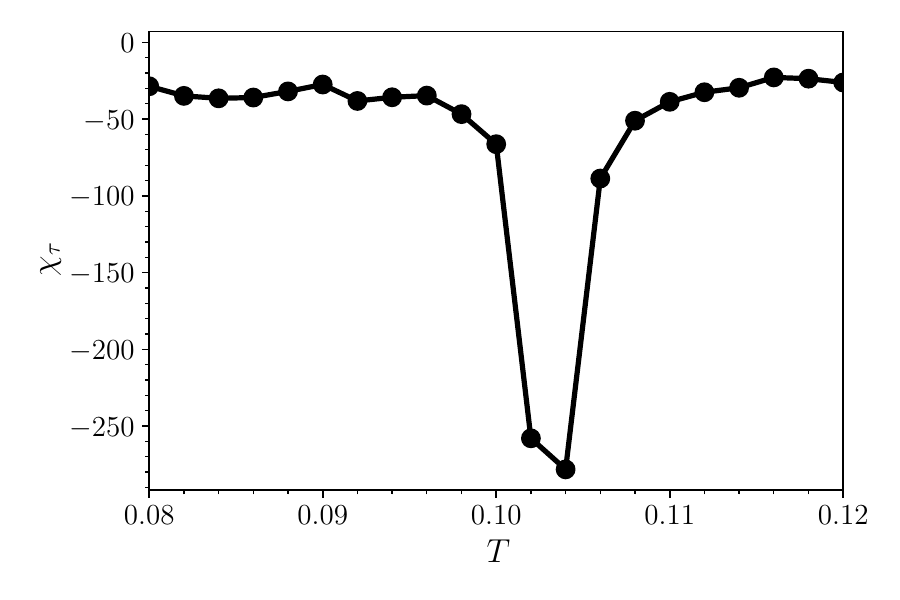}
        \caption{}
    \end{subfigure}
        \begin{subfigure}[b]{0.2\textwidth}
        \includegraphics[width=\textwidth]{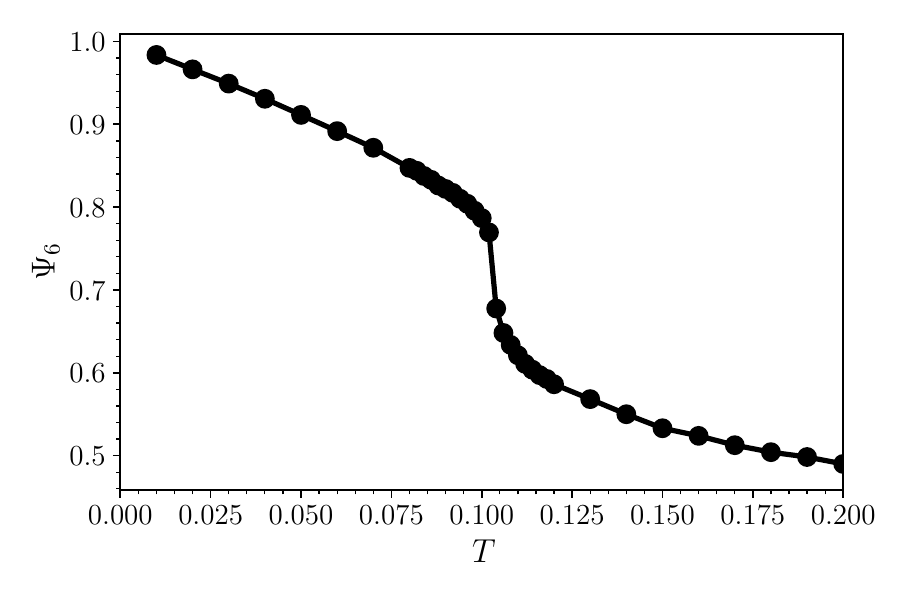}
        \caption{}
    \end{subfigure}
    \begin{subfigure}[b]{0.2\textwidth}
        \includegraphics[width=\textwidth]{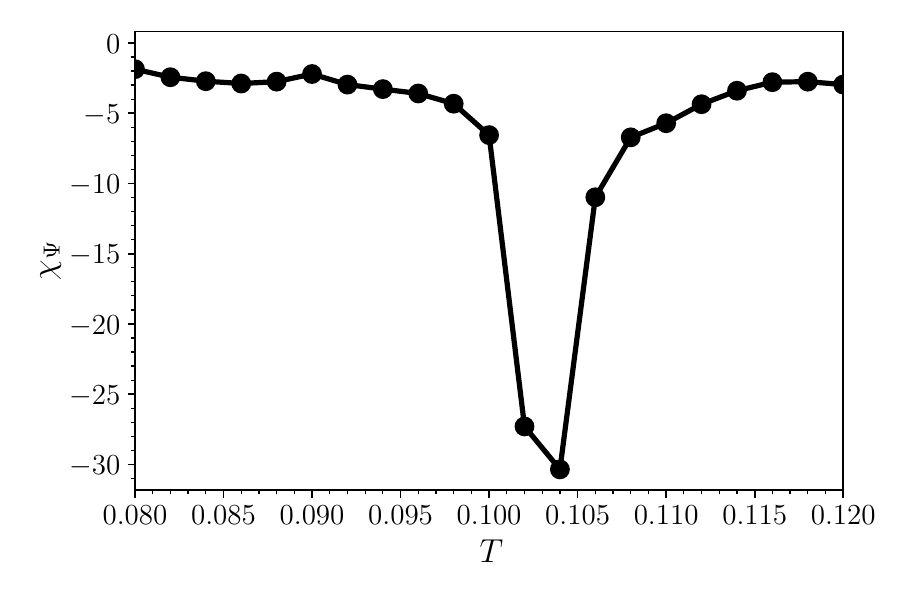}
        \caption{}
    \end{subfigure}
       \begin{subfigure}[b]{0.2\textwidth}
        \includegraphics[width=\textwidth]{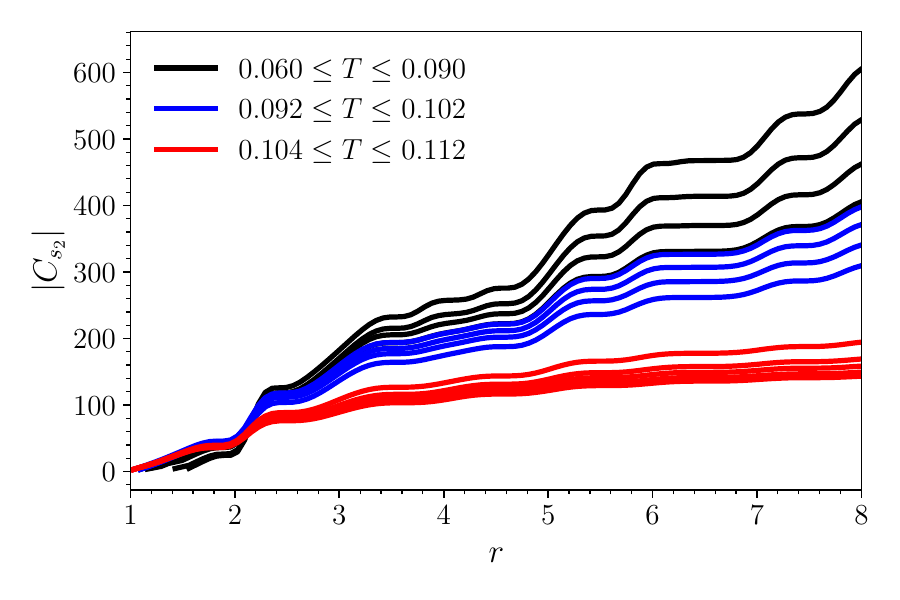}
        \caption{}
    \end{subfigure}
       \begin{subfigure}[b]{0.2\textwidth}
        \includegraphics[width=\textwidth]{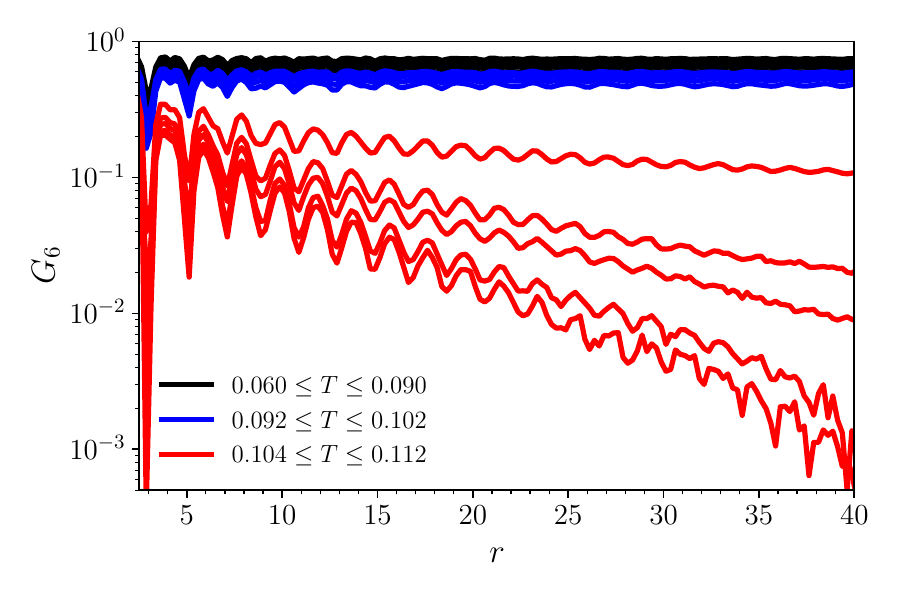}
        \caption{}
    \end{subfigure}
        \begin{subfigure}[b]{0.3\textwidth}
        \includegraphics[width=\textwidth]{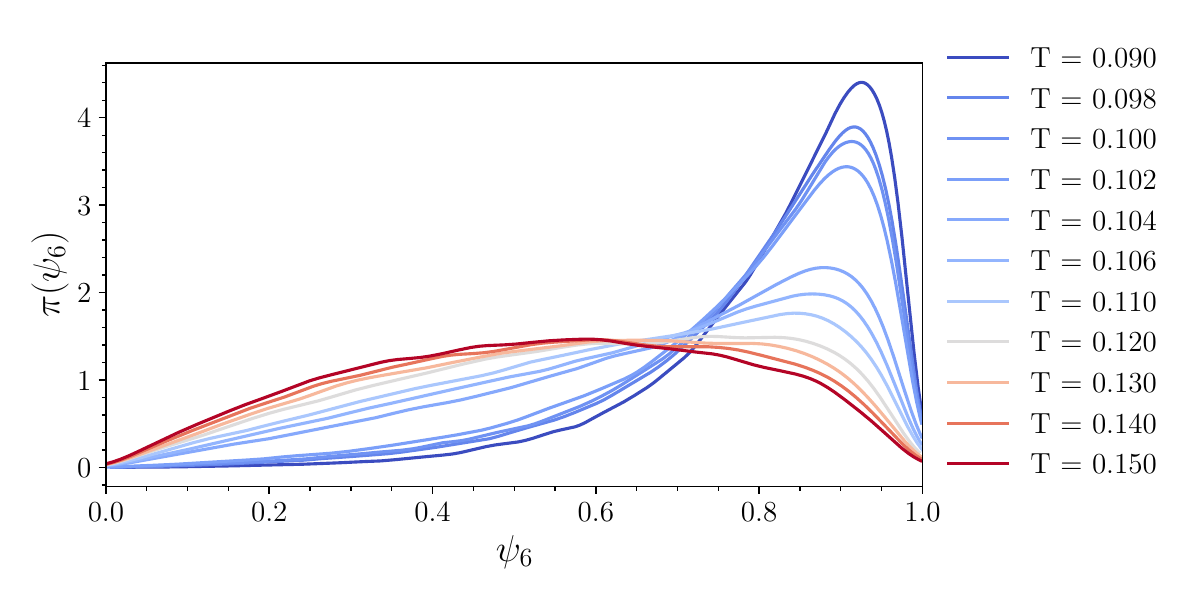}
        \caption{}
    \end{subfigure}
        \caption{ Structural and thermodynamic analysis of the melting of the low-density triangular (LDT) phase.  
(a) Pressure vs. temperature, showing a discontinuous jump at \( T = 0.102 \).  
(b) Radial distribution function \( g(r) \); black: \( T < 0.090 \), blue: \( 0.092 \leq T \leq 0.102 \), red: \( T > 0.102 \).  
(c) Translational order parameter \( \tau \) and (d) its susceptibility \( \chi_\tau \).  
(e) Bond-orientational order \( \psi_6 \) and (f) its susceptibility \( \chi_{\psi} \).  
(g) Pair excess entropy cumulant \( |C_{s_2}| \) and (h) orientational correlation \( G_6(r) \).  
(i) Probability density \( \pi(\psi_6) \).
Error bars are smaller than the symbols.
}
    \label{triangular}
\end{figure}
The system was heated from $T = 0.01$ to $T = 0.200$ to ensure melting in all phases~\cite{Cardoso2021}. The low-density triangular (LDT) phase melts via a first-order transition at \( T = 0.102 \), supported by thermodynamic and structural signatures. The equation of state [Fig.~\ref{triangular}(a)] exhibits a pressure discontinuity consistent with phase coexistence, corroborated by inflection points in the internal energy and sharp peaks in the specific heat - the energy results are presented in  Supplemental Material Section III. The radial distribution function $g(r)$ [Fig.~\ref{triangular}(b)] shows a sharp first-neighbor peak, indicating structural change.

In this phase, particle arrangement is stabilized by the longer of the two characteristic length scales of the core-softened potential: a shoulder at \( r \approx 2.0\sigma \), as opposed to the shorter repulsive core at \( r \approx 1.2\sigma \). Thermal fluctuations enable particles to cross the inter-scale barrier, favoring the shorter length and driving melting via a mechanism distinct from KTHNY theory.

This scenario is further supported by the sharp drop in the translational order parameter \( \tau \) and its susceptibility \( \chi_\tau \) [Figs.~\ref{triangular}(c)--(d)], as well as by discontinuities in the bond-orientational order parameter \( \psi_6 \) and its susceptibility \( \chi_\psi \) [Figs.~\ref{triangular}(e)--(f)].

Additional evidence comes from the pair excess entropy cumulant \( |C_{s_2}| \) and the orientational correlation function \( G_6(r) \) [Figs.~\ref{triangular}(g)--(h)]. Below \( T = 0.102 \), \( |C_{s_2}| \) increases with distance, indicating long-range translational order, while \( G_6(r) \) decays slowly. Above the transition, both quantities saturate or decay exponentially, ruling out the presence of an intermediate hexatic phase. Furthermore, the probability density function \( \pi(\psi_6) \), shown in Fig.~\ref{triangular}(i), reveals an abrupt decrease in the population of particles with high \( \psi_6 \) values as the system crosses the transition, particularly between \( T = 0.102 \) and \( T = 0.104 \). This sharp reduction signals the loss of local bond-orientational order. At higher temperatures (e.g., \( T = 0.150 \)), the fluid becomes fully isotropic, though residual triangular order suggests transient clustering at lower \( T \). Snapshots of the configurations at distinct temperatures are provided in Section II of the Supplementary Material.

Unlike hard-disk systems~\cite{Bernard2011}, which show two-step melting, the LDT phase displays a direct solid–liquid transition with abrupt loss of order. Unlike the KTHNY mechanism~\cite{Kosterlitz1973, Halperin1978, Nelson1979}, melting here arises from a crossover between interaction scales: particles reorganize from long-scale to short-scale configurations, triggering structural frustration and a discontinuous transition.

\begin{figure}[h!]
    \centering
    \begin{subfigure}[b]{0.2\textwidth}
        \includegraphics[width=\textwidth]{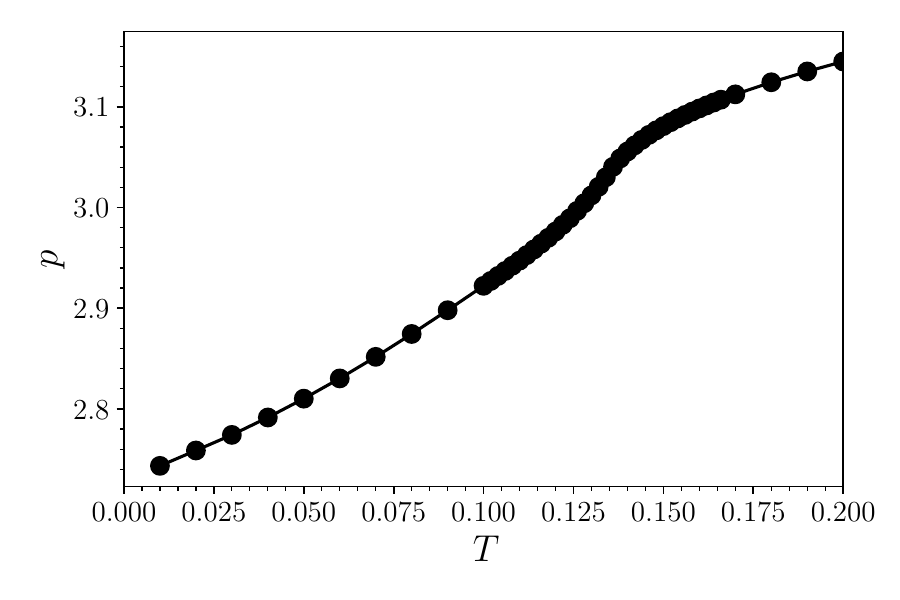}
        \caption{}
    \end{subfigure}
            \begin{subfigure}[b]{0.2\textwidth}
        \includegraphics[width=\textwidth]{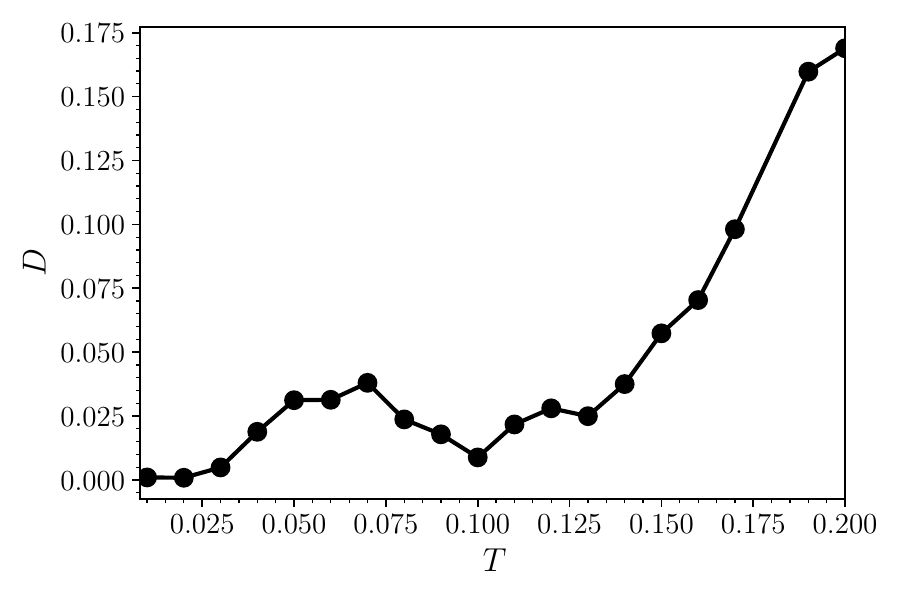}
        \caption{}
    \end{subfigure}
        \begin{subfigure}[b]{0.2\textwidth}
        \includegraphics[width=\textwidth]{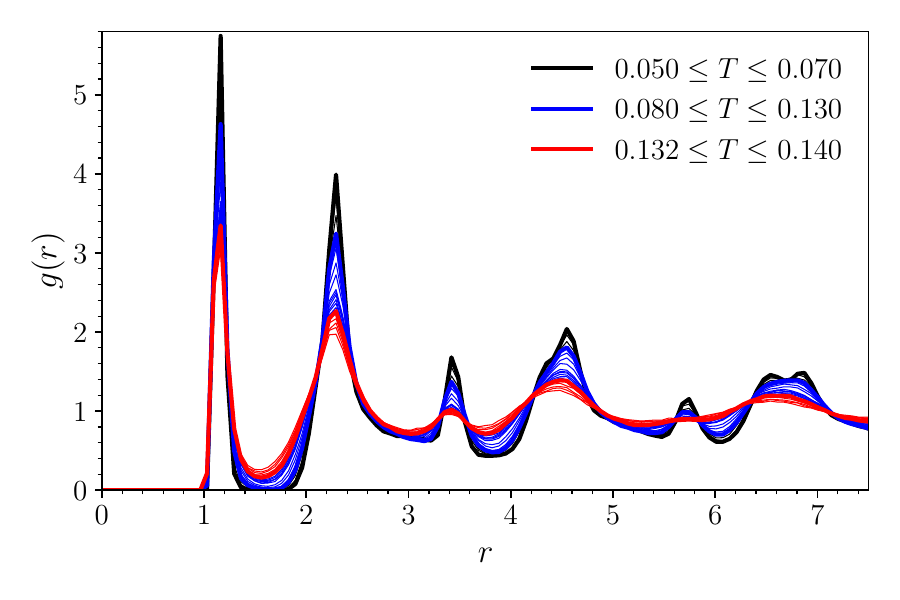}
        \caption{}
    \end{subfigure}   
        \begin{subfigure}[b]{0.2\textwidth}
        \includegraphics[width=\textwidth]{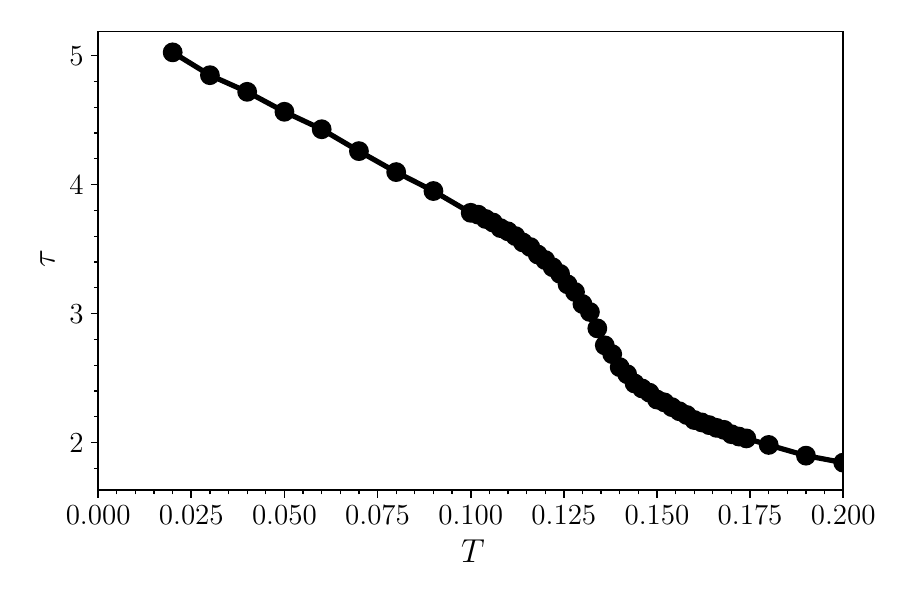}
        \caption{}
    \end{subfigure}        
    \begin{subfigure}[b]{0.2\textwidth}
        \includegraphics[width=\textwidth]{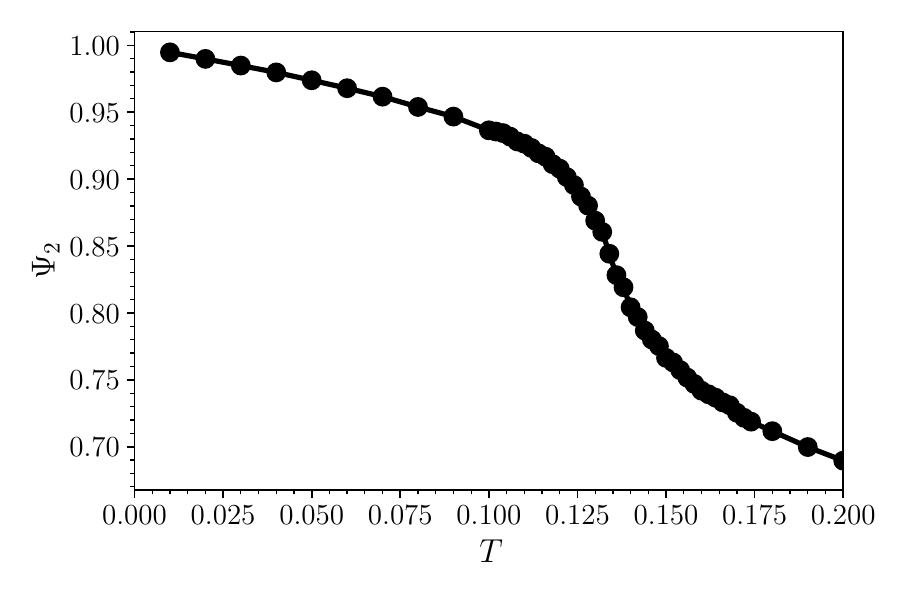}
        \caption{}
    \end{subfigure}    
        \begin{subfigure}[b]{0.2\textwidth}
        \includegraphics[width=\textwidth]{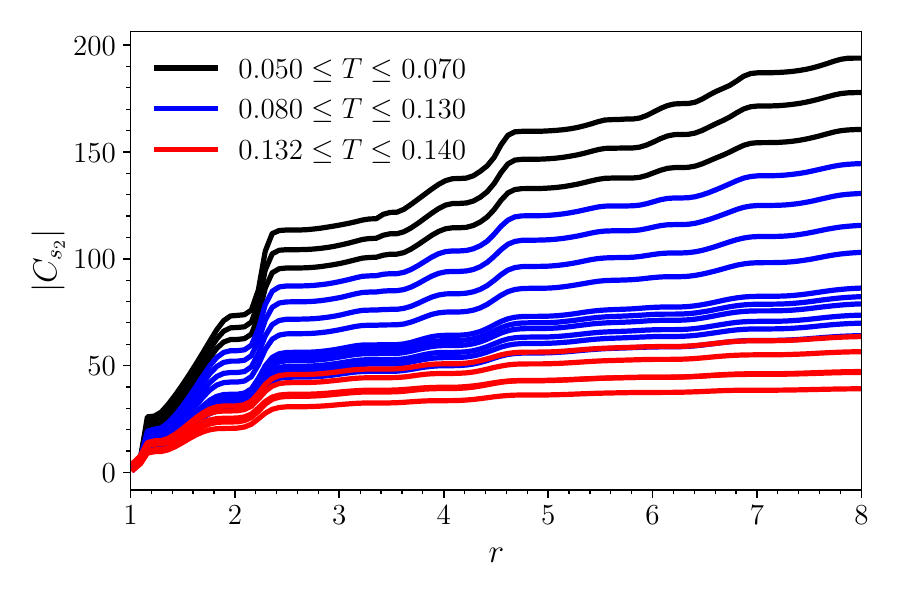}
        \caption{}
    \end{subfigure}
    \begin{subfigure}[b]{0.2\textwidth}
        \includegraphics[width=\textwidth]{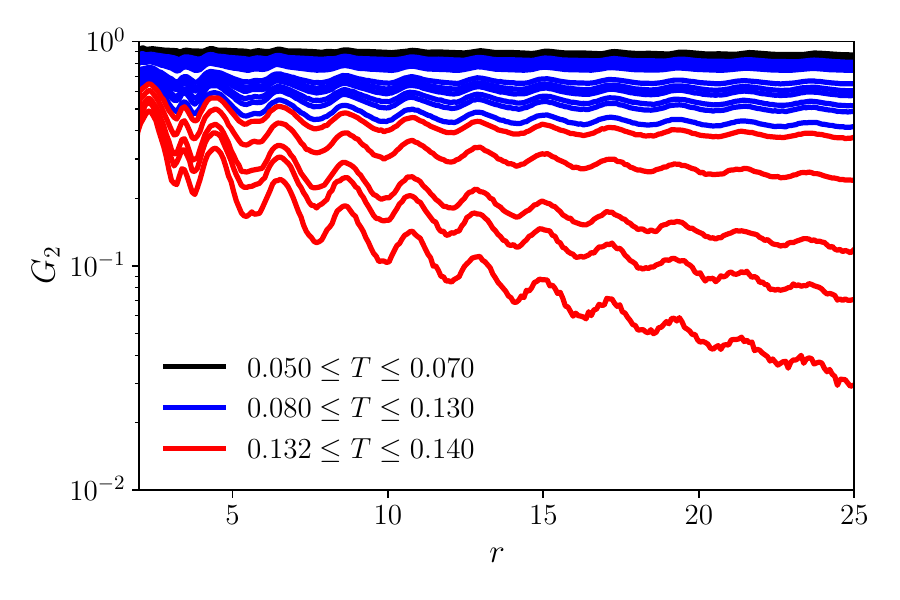}
        \caption{}
    \end{subfigure}
    \begin{subfigure}[b]{0.25\textwidth}
        \includegraphics[width=\textwidth]{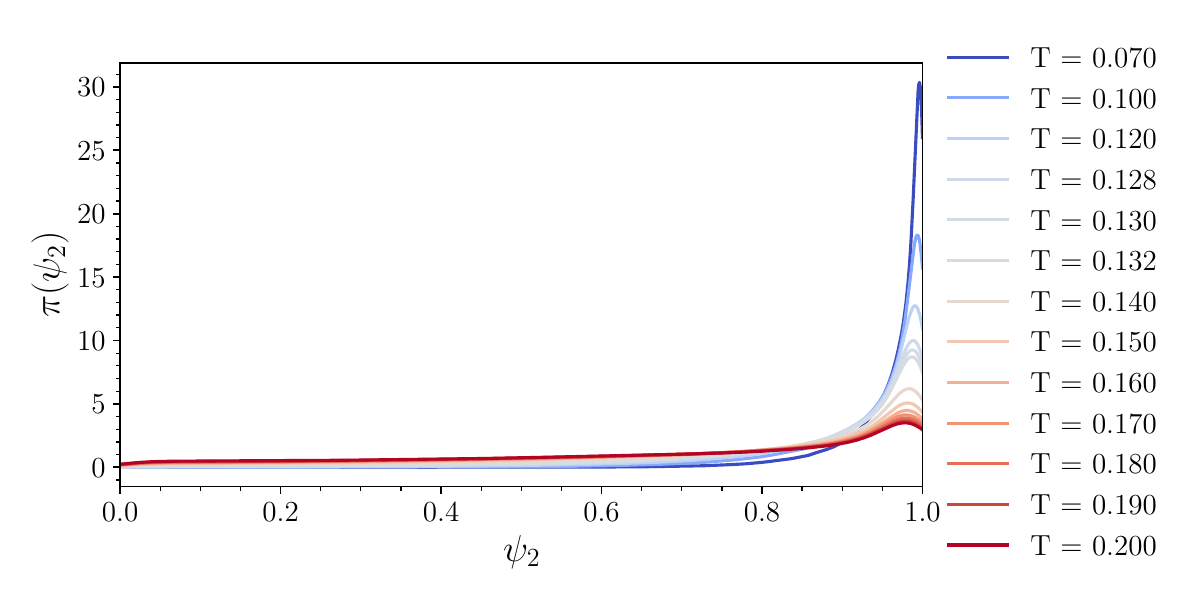}
        \caption{}
    \end{subfigure}
\caption{Thermodynamic, dynamic and structural properties of the stripe phase. (a) Pressure as a function of temperature. (b) Self-diffusion coefficient \( D \). (c) Radial distribution function \( g(r) \); (d) translational order parameter \( \tau \); (e) orientational order parameter \( \Psi_2 \); (f) modulus of the pair excess entropy cumulant \( |C_{s_2}| \); and (g) orientational correlation function \( G_2(r) \). (h) Probability density function \( \pi(\psi_2) \). For panels (c), (f), and (g), the color code is as follows: black curves for \( T = 0.050 - 0.070 \) (ordered stripes), blue for \( T = 0.072 - 0.130 \) (distorted stripes), and red for \( T \geq 0.132 \) (fluid of polymer-like clusters). }
    \label{stripes}
\end{figure}

The stripe phase undergoes a continuous melting transition, in contrast to the first-order behavior observed in the low-density triangular (LDT) phase. The equation of state [Fig.~\ref{stripes}(a)] displays a smooth pressure variation with temperature, without any discontinuities. This is corroborated by the internal energy and constant-volume specific heat \( C_V \) (see Supplemental Material Section III), which show broad, rounded features rather than sharp anomalies, with a maximum centered around \( T = 0.130 \), consistent with continuous thermodynamic behavior.

This melting scenario resembles those reported in systems with competing long-range repulsion and short-range attraction~\cite{Reichhardt2010,Reichhardt2011}, where intermediate structured phases exhibit liquid-like mobility within otherwise ordered domains. Similarly, the self-diffusion coefficient \( D \) in the stripe phase [Fig.~\ref{stripes}(b)] displays non-monotonic behavior: it increases with temperature up to \( T \approx 0.070 \), decreases slightly at intermediate temperatures, and rises again near the melting point. This trend reflects a transition from confined intra-stripe motion to more fluid-like dynamics as thermal fluctuations increase.

Structural analysis based on the radial distribution function \( g(r) \) [Fig.~\ref{stripes}(c)] reveals a gradual loss of positional order. At low temperatures (\( T = 0.050 - 0.070 \)), the stripes are well-aligned and linear; as the temperature rises (\( T = 0.072 - 0.130 \)), the stripes bend and distort, eventually transforming into worm-like polymeric clusters for \( T \geq 0.132 \). This progressive deformation mirrors thermal transitions observed in vortex lattices and colloidal stripe assemblies~\cite{Komendova2013,VlaskoVlasov2014}.

The translational and orientational order parameters \( \tau \) and \( \Psi_2 \), along with their respective correlation functions \( |C_{s_2}| \) and \( G_2(r) \) [Figs.~\ref{stripes}(d)--(g)], decay smoothly with increasing temperature, with no abrupt symmetry breaking. This behavior is consistent with a continuous melting transition involving the unbinding of topological defects, as described by the Kosterlitz-Thouless-Halperin-Nelson-Young (KTHNY) theory~\cite{Kosterlitz1973, Halperin1978, Nelson1979}, although no well-defined intermediate hexatic-like regime is detected.

The microscopic origin of this behavior lies in the anisotropic spatial arrangement imposed by the core-softened potential. The short-range scale \( r_1 \approx 1.2\sigma \) supports intra-stripe mobility, while the longer-range scale \( r_2 \approx 2.0\sigma \) maintains inter-stripe separation. As thermal fluctuations grow, transverse undulations destabilize the stripe configuration, ultimately yielding a structured fluid composed of polymer-like clusters.

This evolution is further illustrated by the probability density function \( \pi(\psi_2) \), shown in Fig.~\ref{stripes}(h). At low temperatures, a sharp peak near \( \psi_2 = 1.0 \) reflects strong local orientational alignment, characteristic of straight stripe configurations. As the stripes bend, the peak broadens and its intensity decreases. Notably, even at high temperatures (\( T = 0.200 \)), a residual peak persists near \( \psi_2 = 1.0 \), indicating that the fluid retains local orientational correlations and a cluster-like character. Representative snapshots of these configurations are provided in Section II of the Supplementary Material.

Overall, the melting of the stripe phase follows a continuous pathway with features reminiscent of liquid-crystalline transitions. The absence of discontinuities and the gradual decay of structural order suggest that the process lies within, or closely adjacent to, the KTHNY universality class. These findings highlight how anisotropy and competing interaction length scales in core-softened systems can stabilize intermediate mesophases and give rise to unconventional melting behavior in two dimensions.

\begin{figure}[h!]
    \centering
    \begin{subfigure}[b]{0.2\textwidth}
        \includegraphics[width=\textwidth]{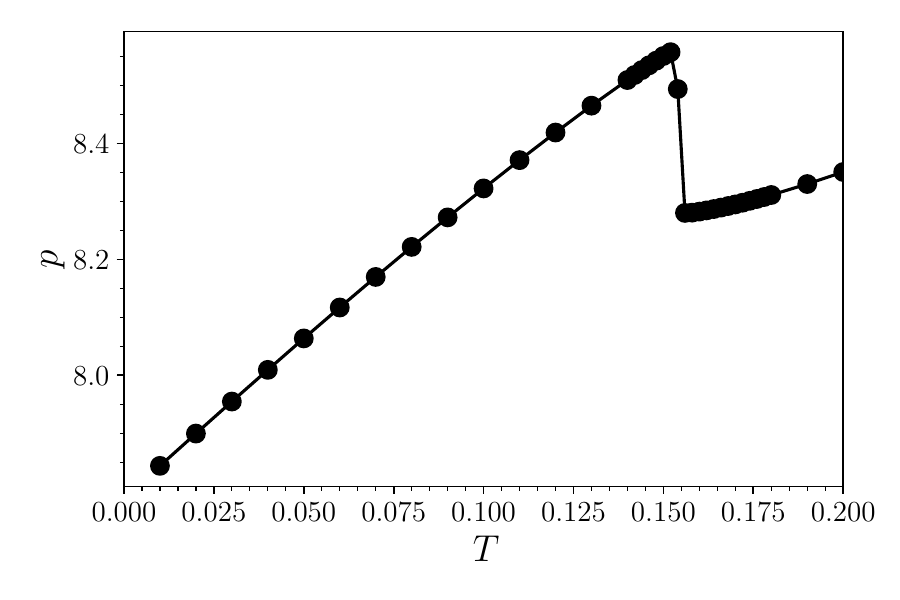}
        \caption{}
    \end{subfigure}
        \begin{subfigure}[b]{0.2\textwidth}
        \includegraphics[width=\textwidth]{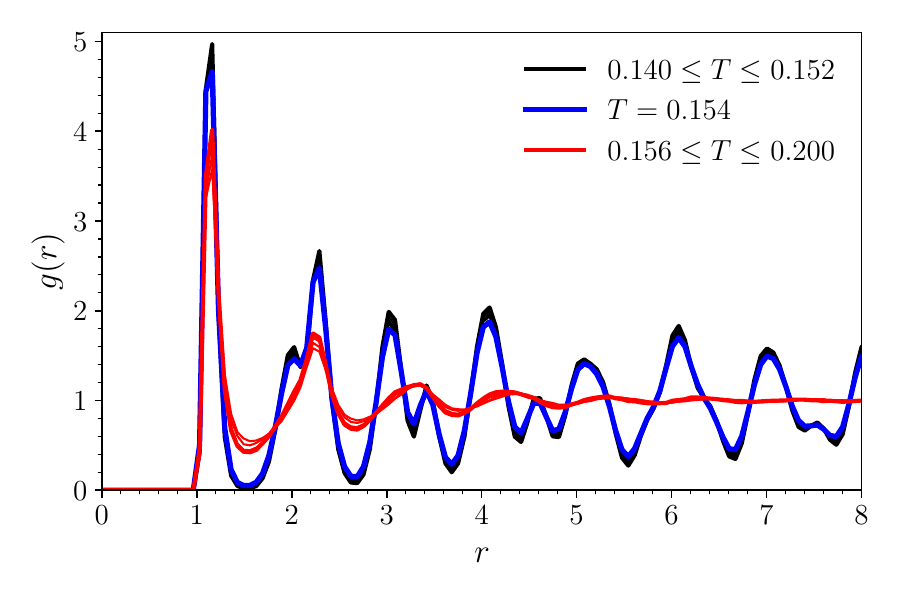}
        \caption{}
    \end{subfigure}   
        \begin{subfigure}[b]{0.2\textwidth}
        \includegraphics[width=\textwidth]{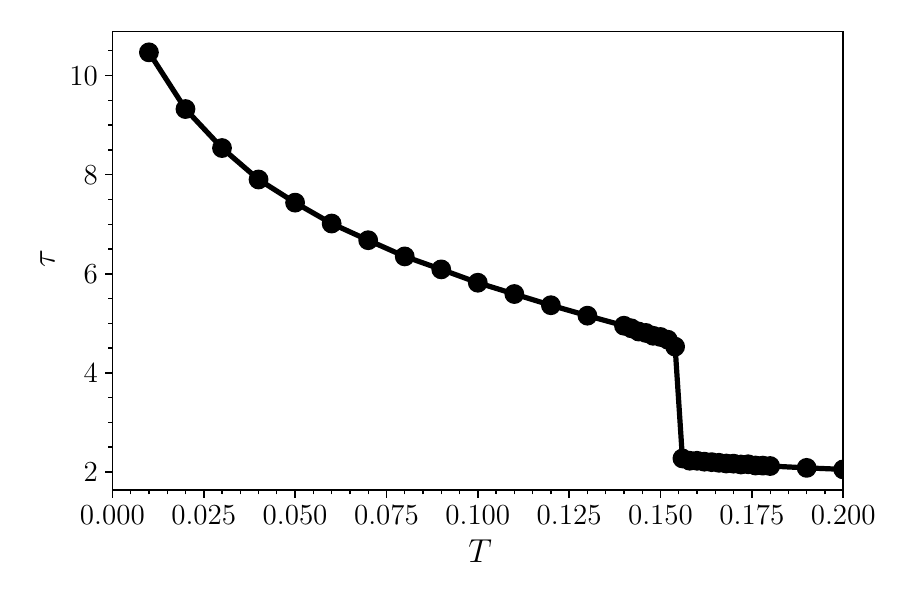}
        \caption{}
    \end{subfigure}  
            \begin{subfigure}[b]{0.2\textwidth}
        \includegraphics[width=\textwidth]{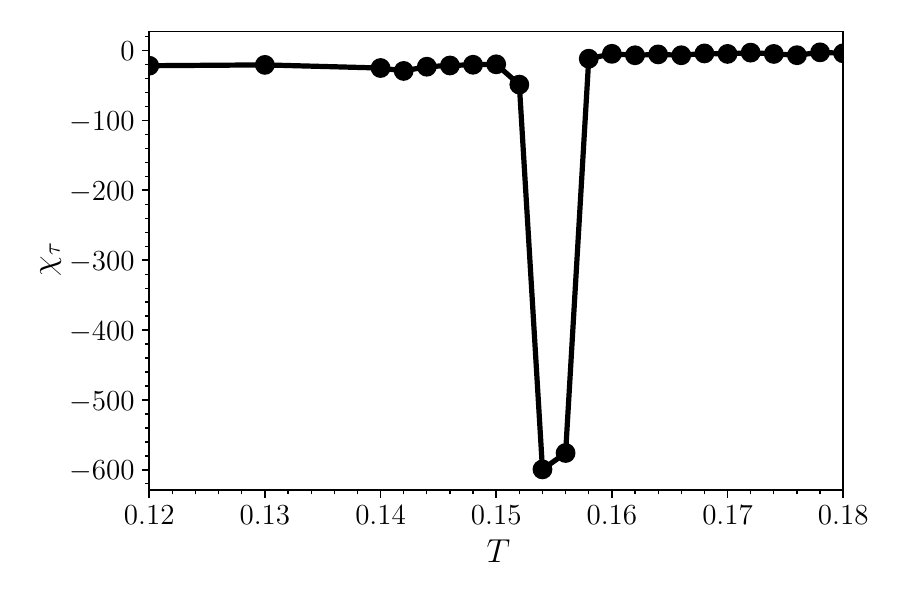}
        \caption{}
    \end{subfigure} 
    \begin{subfigure}[b]{0.2\textwidth}
        \includegraphics[width=\textwidth]{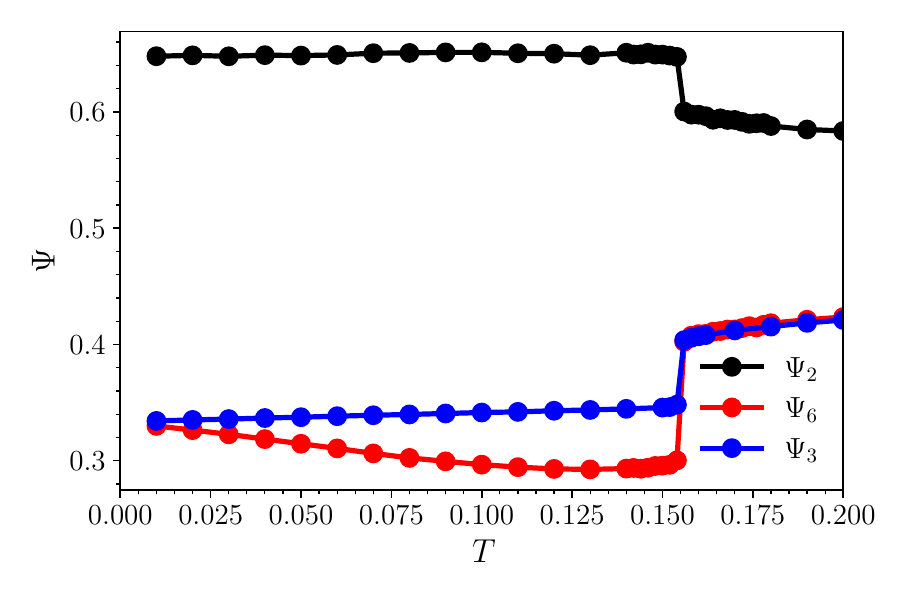}
        \caption{}
    \end{subfigure}
        \begin{subfigure}[b]{0.2\textwidth}
        \includegraphics[width=\textwidth]{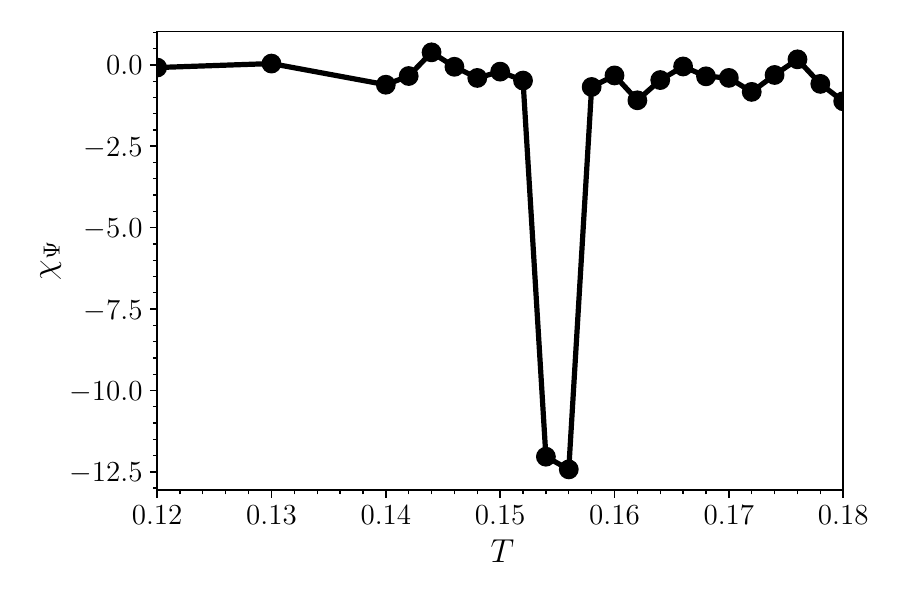}
        \caption{}
    \end{subfigure}
        \begin{subfigure}[b]{0.2\textwidth}
        \includegraphics[width=\textwidth]{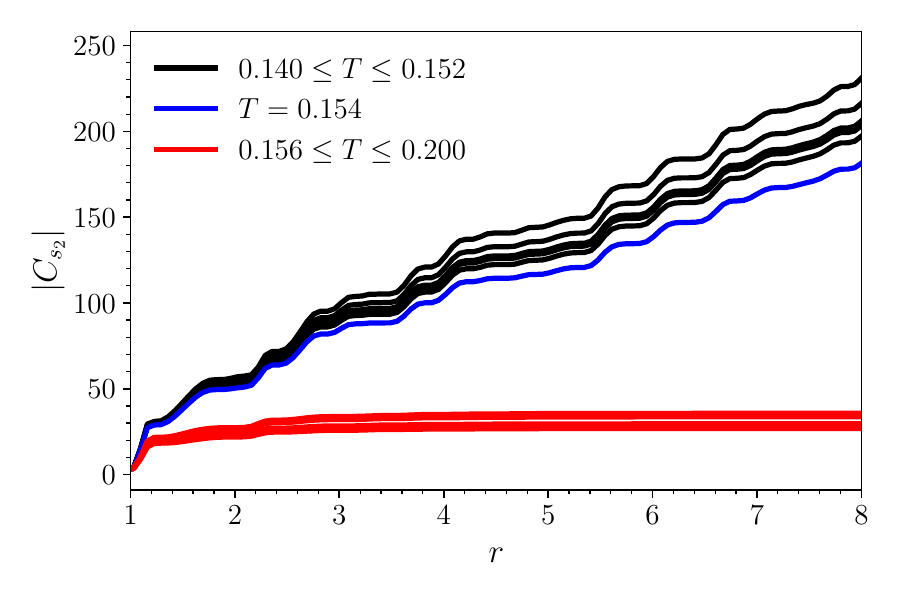}
        \caption{}
    \end{subfigure}
    \begin{subfigure}[b]{0.2\textwidth}
        \includegraphics[width=\textwidth]{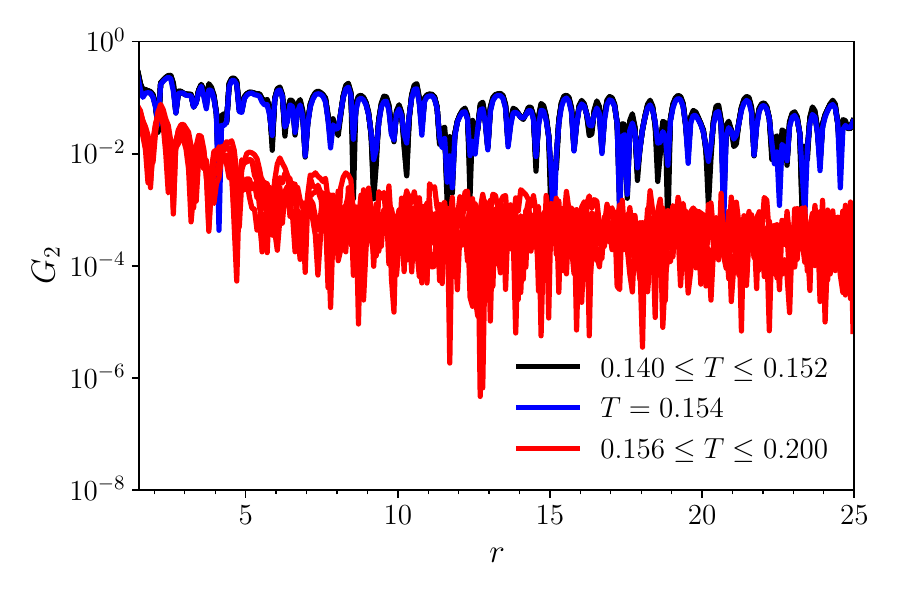}
        \caption{}
    \end{subfigure}
    \begin{subfigure}[b]{0.3\textwidth}
        \includegraphics[width=\textwidth]{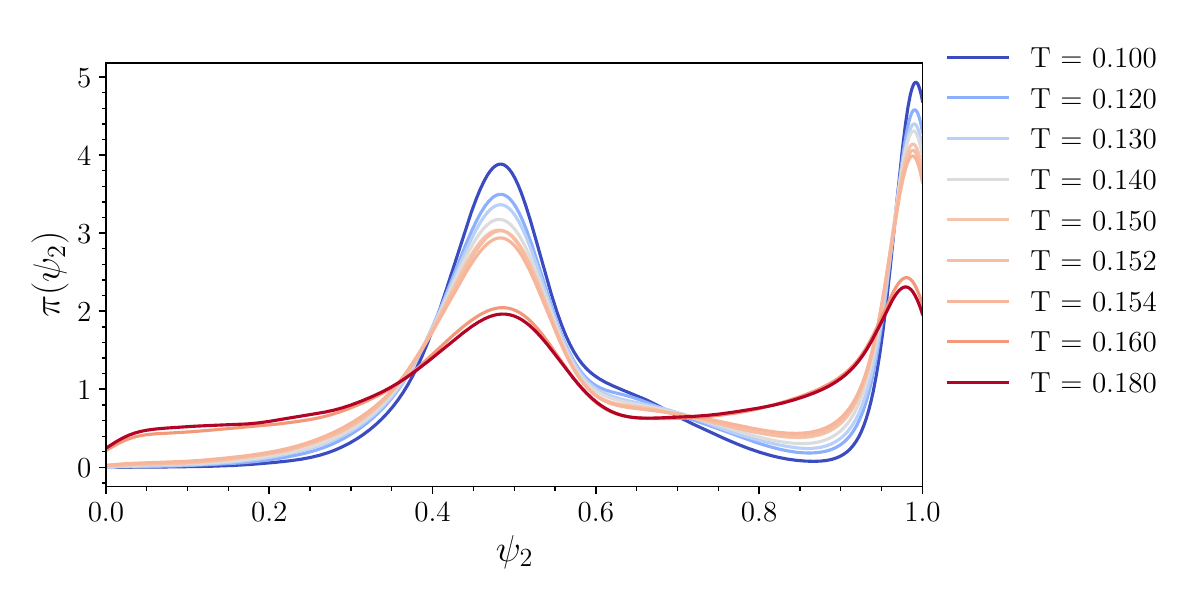}
        \caption{}
    \end{subfigure}
    \caption{Thermodynamic and structural characterization of the kagome phase. (a) Equation of state \( p(T) \). (b) Radial distribution function \( g(r) \), with curves colored as follows: black for \( 0.140 \leq T \leq 0.152 \), blue for \( T = 0.154 \), and red for \( T \geq 0.156 \). (c) Translational order parameter \( \tau \) and (d) its susceptibility \( \chi_\tau \). (e) Orientational order parameters \( \Psi_l \) for \( l = 2 \) (black), \( l = 3 \) (blue), and \( l = 6 \) (red) and (f) Susceptibility \( \chi_{\Psi_2} \). (g) Pair excess entropy cumulant modulus \( |C_{s_2}| \), colored as in (b). (h) Orientational correlation function \( G_2(r) \), also with the same color coding.(i) Probability density function \( \pi(\psi_2) \)}
    \label{kagome}
\end{figure}

The kagome phase undergoes a first-order melting transition that differs qualitatively from those of both the low-density triangular (LDT) and stripe phases. The equation of state [Fig.~\ref{kagome}(a)] exhibits a pressure discontinuity at \( T = 0.154 \). This temperature coincides with an inflection point in the internal energy and a sharp peak in the constant-volume specific heat \( C_V \), shown in Supplemental Material Section III, both hallmarks of a first-order transition~\cite{Bernard2011, loewe2020solid}.

Structural changes across the transition are captured by the radial distribution function \( g(r) \) [Fig.~\ref{kagome}(b)], which shows long-range order for \( T \leq 0.154 \) and becomes featureless for \( T \geq 0.156 \), indicating a transition to an isotropic fluid. This abrupt loss of positional order is mirrored by a sharp drop in the translational order parameter \( \tau \) and a peak in its susceptibility \( \chi_\tau \) [Figs.~\ref{kagome}(c)--(d)].

To probe orientational symmetry, we evaluated \( \Psi_2 \), \( \Psi_3 \), and \( \Psi_6 \), corresponding to 2-, 3-, and 6-fold local symmetries, respectively. As shown in Fig.~\ref{kagome}(e), \( \Psi_2 \) dominates in the ordered phase and decays sharply across the transition, while \( \Psi_3 \) and \( \Psi_6 \) increase slightly in the fluid, reflecting local rearrangements toward isotropic configurations. Similar symmetry-selective responses have been reported in frustrated and deformable particle systems~\cite{loewe2020solid, nishikawa2023liquid}.

A slight offset between translational and orientational indicators is observed: while \( \tau \) and the pair excess entropy cumulant \( |C_{s_2}| \) [Fig.~\ref{kagome}(g)] change abruptly at \( T = 0.154 \), the susceptibility \( \chi_{\Psi_2} \) peaks at \( T = 0.156 \) [Fig.~\ref{kagome}(f)]. However, the orientational correlation function \( G_2(r) \) [Fig.~\ref{kagome}(h)] decays exponentially above the transition, with no indication of algebraic behavior, ruling out the existence of an intermediate hexatic phase~\cite{Kapfer2015}.

Further insight is provided by the distribution of the local bond-orientational order parameter \( \Psi_2 \), which remains bimodal at all temperatures, with peaks near 0.5 and 1.0 [Fig.~\ref{kagome}(i)]. This bimodality reflects the tri-triangular backbone of the kagome lattice, where particles occupy sites with distinct local environments. Even above the melting temperature, the persistence of this distribution—albeit less pronounced—suggests the survival of short-range kagome-like correlations, indicating a clusterized fluid phase. Snapshots of the configurations at distinct temperatures are provided in Section II of the Supplementary Material.

Together, these findings confirm a direct first-order melting transition in the kagome phase, distinct from the continuous pathway observed in the stripe phase and the two-step melting predicted by the KTHNY theory~\cite{Kapfer2015, nishikawa2023liquid}. The suppression of quasi-long-range order and the mismatch between translational and orientational observables underscore the role of geometric frustration and competing symmetries in destabilizing hexatic phases and promoting discontinuous melting.

Our study reveals that melting in two-dimensional colloidal crystals with core-softened interactions is a non-universal process governed by a subtle interplay between lattice symmetry, geometric frustration, and competing interaction length scales. Through a systematic analysis of three polymorphic solid phases—low-density triangular, stripe, and kagome—we demonstrate that each follows a distinct melting pathway. The triangular and kagome phases exhibit abrupt, first-order transitions marked by the simultaneous loss of translational and orientational order, consistent with mechanisms driven by energetic frustration and structural competition. In contrast, the stripe phase undergoes a continuous melting transition reminiscent of the KTHNY scenario, where both types of order decay gradually under thermal fluctuations.

These contrasting behaviors underscore the breakdown of universality in 2D melting and highlight the central role of local symmetry and interaction anisotropy in shaping thermodynamic responses. Beyond advancing our understanding of phase transitions in soft-matter systems with polymorphism and frustration, our findings offer guiding principles for the design of programmable self-assembled materials with tunable phase behavior. In particular, the competition between characteristic interaction length scales emerges as a key mechanism driving abrupt structural transitions, providing a microscopic route to engineer non-universal melting in colloidal assemblies.

{\it Acknowledges --} J.R.B. acknowledges support from CNPq and FAPERGS. A.V.I. was supported by CAPES (Finance Code 001). T.P. acknowledges support from FAPESP. Portions of the manuscript text were refined with the assistance of OpenAI's ChatGPT, which was used to improve grammar, clarity, and consistency

\bibliographystyle{apsrev4-2}
\bibliography{biblio}

\end{document}